\documentclass[11pt]{article}

\usepackage{jjb}
\usepackage{authblk}
\usepackage{xcolor}
\usepackage{tikz}

\graphicspath{{figs/}}

 \usepackage{cite}

\title{The Mathematics of Quantum-Enabled Applications on the D-Wave Quantum Computer}
\author[1]{Jesse J. Berwald}
\affil[1]{D-Wave Systems Inc., 3033 Beta Avenue, Burnaby, British Columbia, Canada V5G 4M9}
\affil[1]{Corresponding author, \href{mailto:jberwald@dwavesys.com}{jberwald@dwavesys.com}}

\date{}

\begin{document}
\maketitle


\section{Introduction}
\label{sec:intro}

Over half a century ago, a ground-breaking technology, the microchip, started appearing in computers and research facilities around the world. Today there is no question of its importance. Yet in 1968, ten years after its invention, it was still a novelty to some: An IBM engineer famously asked, ``But what... is it good for?''\footnote{Attributed to a particularly myopic engineer at the Advanced Computing Systems Division of IBM, 1968, commenting on the microchip.} 
Recent advances in the development of quantum computers in some ways mirror this evolution, though time, experience, and feverish media coverage ensure that few will ask the same naive question. The similarity comes from the observation that quantum computers are on a similar cusp, that of having broad societal impact, as the microchip was in the last century.  

After some reflection, mathematicians and scientists may find themselves asking related questions. For instance, {\em What are quantum computers good for today?}; {\em As a mathematician, what's in it for me?}; and of course, {\em How do they do work?} Other than the last question, there are few definitive answers available. This article attempts to guide the reader towards her own intuition regarding the first two questions, but limits the ``how'' to a cursory glance and a host of references.

This article covers quantum computing from the angle of {\em adiabatic quantum computing}~\cite{Kadowaki1998,Farhi2001,Albash2018a}, which has proven to have the shortest horizon to real-world applications, partly due to a slightly easier path to development\footnote{But by no means trivial, as D-Wave's 19-year journey can attest to.} than alternative approaches such as {\em gate-model} quantum computers. 

In this article we cover the canonical problem formulation necessary to program the D-Wave quantum processing unit (QPU) and discuss how such a problem is compiled onto the QPU. We also cover recent joint work solving a problem from topological data analysis on the D-Wave quantum computer. The goal of the article is to cover the above from a mathematical viewpoint accessible to a wide range of levels to introduce as many people as possible to a small portion of the mathematics encountered in this industry.




\section{Quantum computing background}
\label{sec:bkgd}

\subsection{Historical background}
\label{sec:history}

Richard Feynman is credited with the initial ideas for computing with quantum mechanics, presented in a seminal talk and subsequent paper from 1982, title {\em Simulating Physics with Computers}~\cite{Feynman1982}. 
Significant progress over the past decade has brought the quantum computer industry into what some term the {\em noisy intermediate-scale quantum} (NISQ) era~\cite{Preskill2018}. While quantum computers have yet to show an undeniable advantage over classical systems, their theoretical advantages are well-documented. Quantum annealing, the model adopted by D-Wave, also promises quantum speedup \cite{Somma2012}.
Particularly noteworthy are Shor's algorithm \cite{Shor1995} and Grover search algorithm \cite{Grover1997} for gate-model quantum computers.  Already, in a number of narrowly-defined use cases, improvements over classical computers have been observed on the D-Wave quantum computer~\cite{Denchev2016,King2017,Mott2017,Li2018a,Albash2018}.

\subsection{Technical background}
\label{sec:tech}

The D-Wave quantum computer is a programmable quantum computer whose QPU is comprised of a grid of
superconducting loops, each of which acts as a programmable flux qubit \cite{Harris2010a}. Each element of the grid is a bipartite graph, also known as a Chimera unit cell. 

The loops may have current in either direction, corresponding to
up and down spins. The individual qubits are arranged in a grid
with couplers corresponding to controllable mutual inductances
between the magnetic fields associated with the current loops.

\begin{figure}
	\centering
	\includegraphics[width=0.7\textwidth]{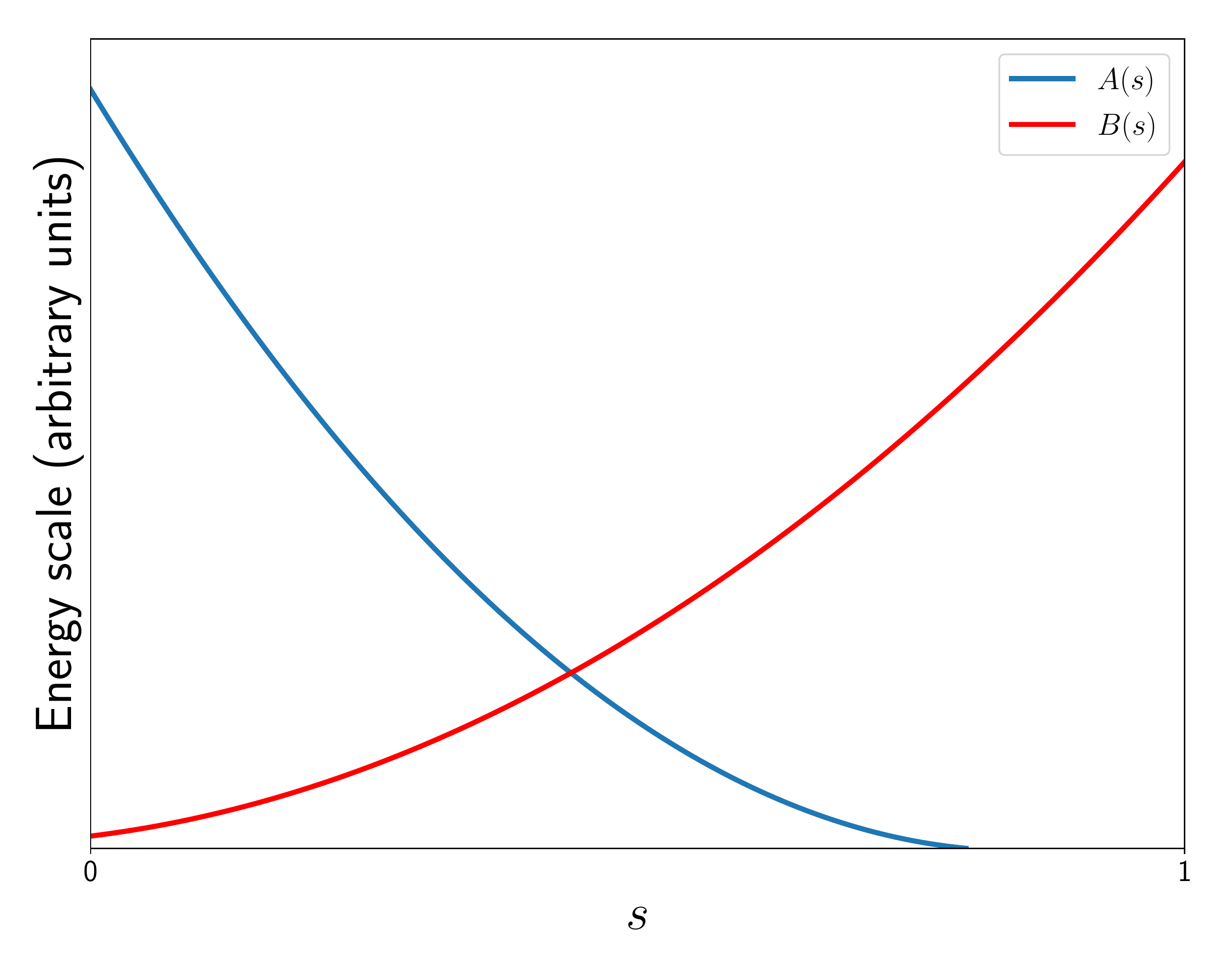}
	\caption{A typical annealing schedule.  Scales are arbitrary; energy is measured in GHz. The parameter $s$ evolves from 0 to 1, on the scale of microseconds. The causes the system to evolve from a purely quantum state in which $A(0) \gg B(0)$, to a classical state where $B(1) \gg A(1)$.}
	\label{fig:annealing}
\end{figure}

The D-Wave quantum computer implements a process known as quantum annealing~\cite{Kadowaki1998,Farhi2001}. 
The goal of a quantum annealing computer is to find a low-energy state of a {\em problem Hamiltonian}, $H_P$, by first passing through a quantum state.
The state evolution is governed by the time-dependent Schr{\"o}dinger equation~\cite{Farhi2001} of the form
\begin{align}\label{eq:formal-ham}
H_0(s) =& A(s) H_D + B(s) H_P  \\ 
   H_D =& - \sum_{i}\sigma_{i}^x \label{eq:ham1} \\ 
   H_P =& \sum_{i} h_i \sigma_{i}^z + \sum_{i,j} J_{i,j}\sigma_{i}^z \sigma_{j}^z, \label{eq:ham2} 
\end{align}
where the $\sigma^x$ and $\sigma^z$ are spin variables taking values in $\{\pm 1\}$~\cite{Boixo2016}. Typical curves, $A(s)$ and $B(s)$, governing the annealing schedule of the system are shown in \cref{fig:annealing}. 
The Hamiltonian, $H_D$, defines the transverse field. The structure of $H_P$
is defined physically by local, real-valued fields $h_i$ and $J_{i,j}$. The system is evolved from an initial quantum ground state, $s=0$, in which the transverse field is high and all possible states are in superposition, to a final classical state at $s=1$. 
The adiabatic theorem of Born and Fock~\cite{Born1928} guarantees that if the time evolution of the system is slow enough then the system will remain in its ground state. Thus, at the end of a slow anneal process, from $s=0$ to $s=1$, the ground state from the quantum state will also be the global minimum of the problem Hamiltonian, $H_P$. 
We are not concerned here with the energy scale, so in the algebraic case it suffices to define $H_0$ using the single parameter $\Gamma(s) = \frac{A(s)}{B(s)}$, leading to
\begin{align}\label{eq:gammaH}
	H_0(s) = \Gamma(s) H_D + H_P.
\end{align}
With this simplification, annealing begins with the annealing parameter $\Gamma \gg 1$ in the quantum state, and ends in the classical state with $\Gamma \ll 1$.


Precise control over the transverse field and the configuration of the spin variables, as exists on the D-Wave quantum computer, allows for novel quantum material simulation \cite{Harris2018}, one of the central ideas in Feynman's original paper. From a more mathematical perspective, we note that during the annealing process the D-Wave quantum computer samples from an approximate Boltzmann distribution over the energy landscape defined by $H_0$. Hence, it is possible in a single run to return many thousands of samples. We discuss this in more detail in \cref{sec:samples}. 

\subsection{Mathematical definition of the problem -- a polynomial viewpoint}
\label{sec:problem-def}

We now divorce ourselves from the physics and focus solely on the problem Hamiltonian going forward. The D-Wave quantum computer is designed to seek a minimum energy solution to an Ising problem, \cref{eq:ham2}, or equivalently a combinatorial optimization problem known as a {\em quadratic unconstrained binary optimization} (QUBO) problem. 
A simple transformation takes the Ising formulation to a QUBO using $x \mapsto \frac{x + 1}{2}$. 

We now describe the QUBO formulation of the problem Hamiltonian in more detail. We begin with a couple of definitions. 
\begin{dfn}\label{def:binary}
	Let $\B^n := \Z_2^{n}$, the $n$-dimensional hypercube of binary vectors. 
\end{dfn}

After a transformation of variables, we can define the problem Hamiltonian as a QUBO taking arguments from $\B^n$.

\begin{dfn}\label{def:Hqubo}
	Let $H_P(\x) := \sum_{i=1}^n h_i x_i + \sum_{i=1}^n \sum_{j=1}^n J_{i,j} x_i x_j$ be a real-valued polynomial with arguments $\x \in \B^n$.
\end{dfn}

The polynomial $H_P$ is an interesting mathematical object: The coefficients of $H_P$ live in $\Q$, yet the variables are restricted to $\B^n$. To deal with this, at least notationally, we define a restriction to the polynomials over the rationals.\footnote{We could also take coefficients over $\R$.} 
\begin{dfn}\label{def:Qx}
	Define the set of polynomials with rational coefficient and binary variables, $\x = (x_1,\ldots, x_n) \in \B^n$, as $\Q[\x|\B^n] \subset \Q[\x]$. 
\end{dfn}

\begin{rmk}
	We can regard $\Q[\x|\B^n]$ as a quotient ring, $\Q[x_1,\ldots, x_n] / \langle{x_1^2 - x_1, \ldots, x_n^2 - x_n}\rangle$. The ideal $\langle{x_1^2 - x_1, \ldots, x_n^2 -x_n}\rangle$ absorbs all polynomials in $\Q[\x]$ for which the binary constraint, $x_i^2 = x_i$, holds. 
\end{rmk}

This remark points to an elegant area of research. Dridi \etal.~\cite{Dridi2017} leverage the algebraic properties of $H_P$ in developing a number of applications using the D-Wave quantum computer. For instance, in \cite{Dridi2017} they leverage Groebner bases to reduce the size of the problem prior to sending it to the quantum computer. In \cite{Dridi2018}, Dridi \etal., leverage computational algebraic geometry for the important problem of embedding the problem Hamiltonian onto the QPU.

With \cref{def:binary} and \cref{def:Qx}, we can now state the problem solved by the quantum computer more precisely. 
\begin{dfn}\label{def:minH}
	Suppose we are given a Hamiltonian $H_P \in \Q[\x | \B^n]$ and a quantum computer, $\QQ$, designed to implement the adiabatic theorem using the time-dependent Schr{\"o}dinger equation \cref{eq:formal-ham}. Then $\QQ$ solves the combinatorial optimization problem
	\begin{align}\label{eq:minH}
		\HH \equiv \argmin_{\x \in \B^n} H_P(\x),
	\end{align}
	given proper assumptions on the evolution of the system $H_0$.
\end{dfn}
The combinatorial optimization problem defined by $\HH$ represents abstractly the problem to be solved. These are, in general, NP-hard problems, making the prospect of a quantum annealing computer which can solve the class of problems described by $\HH$ enticing. Lucas~\cite{Lucas2014} provides a nice overview of methods for formulating a number of NP-hard problems as QUBOs.

\subsection{Local searches using reverse annealing}
\label{sec:ra}

The previous section defines what is known as forward annealing, which amounts to a global search problem. In this case, $\QQ$ is a black box which takes as input $H_P$ and returns $\x^*$, the minimizer of $\HH$. It is possible to modify $\QQ$ to enable specifying some initial condition for the state as well as the problem description, $H_P$. The ability to search $\B^n$ locally becomes possible in this context.

Recently D-Wave introduced a feature known as {\em reverse annealing} that allows $\QQ$ to accept $H_P$, as well as an initial state from which to anneal. As above, local refers to Hamming distance in $\B^n$. Reverse annealing allows quantum refinement of solutions from a classical starting point. Given a specific state $\x^*$, the D-Wave quantum computer initiates a local search from $\x^*$ by annealing backward from the classical state. Starting with $\Gamma \ll 1$, the quantum computer then increases the transverse field to a mid-anneal quantum superposition, which is equivalent to increasing the annealing parameter to $\Gamma^* \gg 1$. It is possible to pause at $\Gamma^*$ to allow the system to search the space. The system then quenches the transverse field to return to a new classical solution by decreasing annealing parameter to $\Gamma \ll 1$. 
Recent work uses this feature to probe exotic states of matter on the D-Wave quantum computer~\cite{King2018}.

Reverse annealing fits well within hybrid algorithms, especially evolutionary algorithms and heuristic search packages \cite{Chancellor2017}. These algorithms often quickly find good local minima, but can be hampered by local energy barriers. The D-Wave quantum computer is effective at tunneling past tall, thin energy barriers~\cite{Denchev2016}. 

\subsection{Compiling $H_p$ -- a graph minor embedding problem}
\label{sec:embedding}

 Much like a classical computer converts high-level, abstract, and human-readable languages to machine instructions, \cref{eq:minH} must be converted to a {\em quantum machine instruction} (QMI) that will run on the quantum computer. There are numerous steps in this process, one of which, {\em embedding}, we touch on briefly in this section. It is convenient to view the problem Hamiltonian, $H_P$, as a weighted graph. Define $G = \langle{V,E}\rangle$, where the node set
\begin{align*}
	V = \{(u_1,\alpha_1),\ldots, (u_n, \alpha_n) \mid  \alpha_i = h_i\}
\end{align*}
is composed of nodes $u_i$ that are in direct correspondence with each binary variable $x_i$, where $\x= (x_1,\ldots,x_n) \in \B^n$. 
Each node is weighted by the bias $\alpha_i = h_i$. Similarly, the edge set is composed of weighted edges defined by the coupling terms in $H_P$, so that
\begin{align*}
	E = \{(u_i,v_j,\omega_{i,j}) \mid u_i,v_j \in V \text{ and } \omega_{i,j} = J_{i,j},\, J_{i,j} \ne 0\}.
\end{align*}
This definition of $E$ encodes the variable coupling in $H_P$.

The graph $G$ must be embedded onto the hardware to solve $H_P$. 
Embedding the {\em logical graph}, $G$, onto the {\em hardware graph}, $C$, amounts to finding a {\em minor embedding}. A {\em minor} of a graph $H$ is a subgraph of $H$ obtained by contracting or deleting edges, or omitting isolated vertices. A minor embedding is a function that maps the vertices of $G$ to the power set of the vertices of $C$,
\begin{align*}
	\psi : V_G \rightarrow 2^{V_C},
\end{align*}
such that for each $u \in V_G$, the image $\psi(u) \in V_C$ is connected. These connected components within the hardware graph are termed {\em chains}. Embeddings for which $\psi(u)$ is a singleton for all $u$ are called {\em native} embeddings. Lastly, there exists an edge between $\psi(u)$ and $\psi(v)$ whenever $u$ and $v$ are adjacent in the logical graph. The map $\psi$ is the  minor embedding we seek. Whether $G$ can be embedded as a minor contained in $C$ is known to be NP-hard. Elegant, polynomial-time algorithms exist for embedding problem Hamiltonians onto the the QPU architecture~\cite{Cai2014,Boothby2015}. The more efficient the embedding -- the shorter the chains -- the larger the problem that can be solved on the QPU.



\section{A mathematical application}
\label{sec:tda}



Many users of the D-Wave quantum computer in recent years have focused on hybrid workflows. In the context of quantum computing, these are software pipelines that use classical computers for a majority of their work, inserting quantum computation at compute-intensive bottlenecks. 
This is a fruitful area to focus research efforts as there will always be vast amounts of pre- and post-processing within real-world pipelines. Much of that processing is not amenable to quantum acceleration, yet alleviating bottlenecks has the potential to yield significant computational gains.

\begin{figure}[t]
	\centering
	\includegraphics[width=0.8\textwidth]{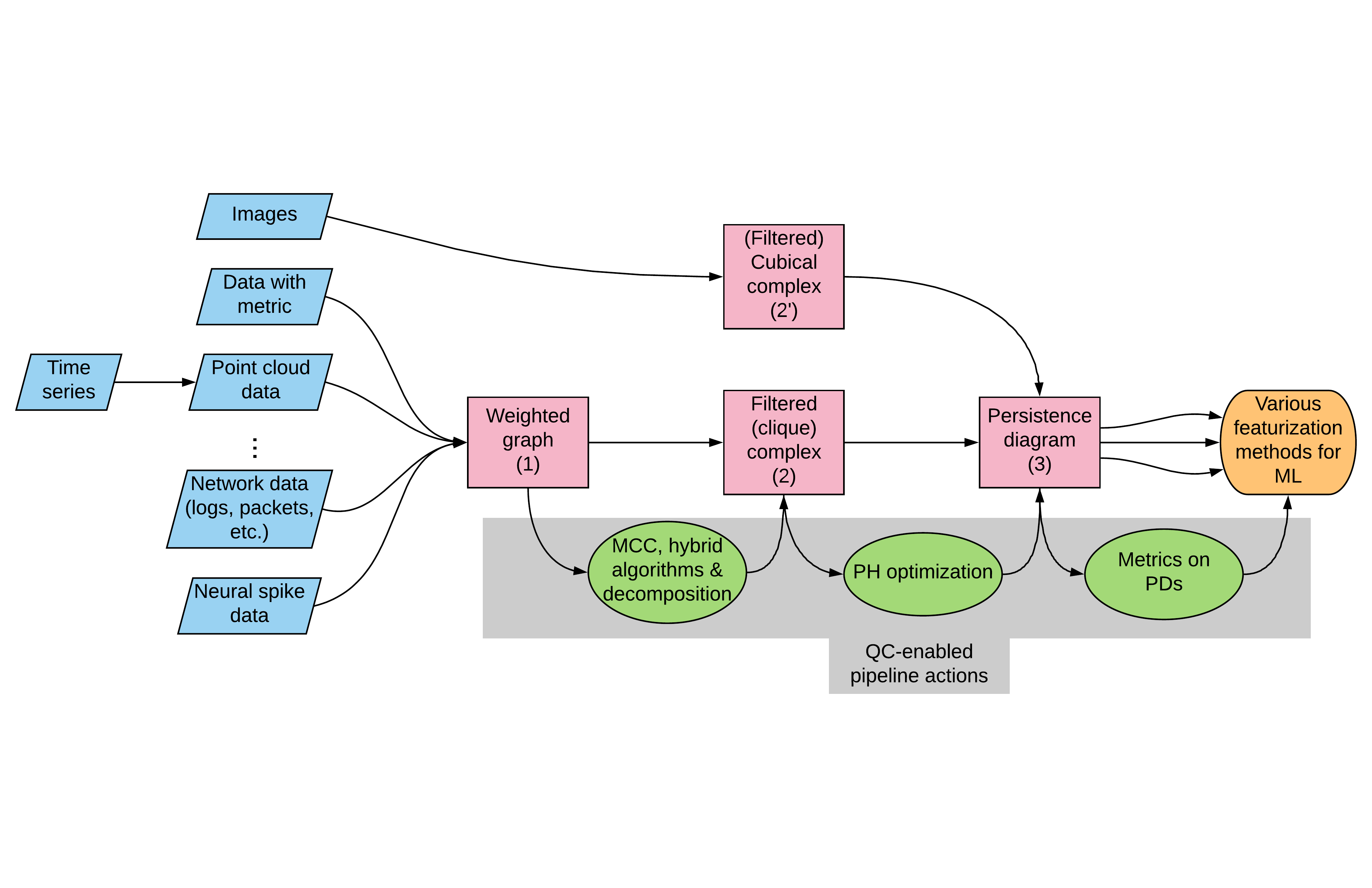}
	\caption{The topological data analysis pipeline for persistent homology. Transformations containing portions amenable to quantum computation are highlighted in the green ovals.}
	\label{fig:tda-pipeline}
\end{figure}

In general, users seeking quantum speedup tend to isolate the tight ``inner loop'' of their problem, the bottleneck where computing the entire loop in one step will reduce the complexity of the problem by orders of magnitude. 
In \cref{fig:tda-pipeline}, we show a typical topological data analysis (TDA) pipeline. The blue boxes on the left represent various potential data sources, while the red boxes in the middle, labeled 1, 2, 2', and 3, show computational bottlenecks in the TDA pipeline. The green ovals highlight the algorithms that could potentially run on the quantum computer to alleviate bottlenecks. Lastly, the final box on the right assumes further processing using the features extracted using the TDA pipeline.

In the next two subsections we provide a brief summary of persistent homology and Wasserstein distance. In \cref{sec:wg-qubo}, we translate the Wasserstein distance to a QUBO to compare the topological signature of point clouds. We do not claim a speed improvement over current state-of-the-art algorithms. Rather, as an example of what is possible, our work shows the interplay between the underlying mathematics of the Wasserstein distance and the construction of an optimal QUBO to solve the combinatorial optimization problem in \cref{def:minH}. 

\subsection{Persistent homology}
\label{sec:pers-hom}

Persistent homology \cite{Edelsbrunner2002,Zomorodian2004,Oudot2017a}, one of the most widely used tools in the field of TDA, is based on the idea that analyzing noisy data using a sequence of resolutions enables one to robustly identify and quantify structure in such data. We summarize this concept below. 

Considering the data at various resolutions yields a filtered topological space built from the data, such as a simplicial or cubical complex for computational purposes,
\begin{equation*}
\emptyset \subseteq K_1 \subseteq K_2 \subseteq \cdots \subseteq K_n = K.
\end{equation*}
Functoriality of homology yields a {\em persistence module}
\begin{equation*}
  \begin{tikzcd}
	0 \ar[r] & H_*(K_1) \ar[r] & H_*(K_2)  \ar[r] & \cdots \ar[r] & H_*(K_n) = H_*(K),
  \end{tikzcd}
\end{equation*}
where homology is computed over a field, $\kappa$, which is often taken to be $\Z_2$. 
The sequence of vector spaces and linear transformations above are studied to understand the structure of the filtration, which serves as a proxy for the structure of the original data.
For computational purposes, a discrete persistence module $\VV$ is a collection of vector spaces and linear transformations of the form
\begin{equation*}
  \begin{tikzcd}
\VV = ( V_1 \ar[r,"\phi_1"] & V_2 \ar[r, "\phi_2"] & \cdots \ar[r, "\phi_{n-1}"] &V_n),
  \end{tikzcd}
\end{equation*}
where the linear transformations $\phi_i^j$, for $i<j$, are given by composition $\phi_{j-1} \phi_{j-2} \cdots\phi_i$.
The interval module $I_{[a,b)}$ is a key object in persistent homology theory that defines the decomposition of the persistence module. For each vector space in $\VV$, let
\begin{equation*}
  V_i =
\begin{cases}
\kappa   & i \in [a,b)\\
0 & \text{else}
\end{cases}
\qquad
\phi_i^j =
\begin{cases}
id   & a \leq i \leq j < b\\
0 & \text{else}.
\end{cases}
\end{equation*}
With reasonable assumptions on the structure of $\VV$~\cite{Bubenik2018}, a persistence module can be decomposed into interval modules
\begin{equation*}
  \VV \cong \bigoplus_{[a,b) \in \BB} I_{[a,b)}.
\end{equation*}
The collection $\BB$ is unique, even though decomposition may not be. 

To visualize $\BB$ we construct a {\em persistence diagram}. Each interval $[a,b)$ is considered as a point $(a,b) \in \R^2$. A point $(a,b)$ in a persistence diagram representing a feature in the data is {\em born} at $a$ and {\em dies} at $b$, and has a lifetime of $b-a$.  
Since $a < b$, we include the diagonal $\Delta = \{(c,c) \mid c \in \R\}$ when drawing a persistence diagram.

The intuition underlying persistent homology is a point in the persistence diagram far from the diagonal represents a homology class that appeared early in the filtration and died late. 
Thus, such a homology class represents a robust topological feature within the noisy data. See  \cite{Edelsbrunner2010} and \cite{Munch2017} for more details. 

\subsection{Wasserstein distance}
\label{sec:wass}

In full generality, a persistence diagram is a finite multiset of points in the plane. Define the region in the plane occupied by points in the persistence diagram as $\R_{\Delta}^2 := \{(c,d) \mid d > c \text{ and } c \ge 0 \} \subset \R^2$. For technical reasons, each diagram includes an additional set of countably infinite copies of each point on the diagonal, $\Delta := \{(d,d) \mid d \ge 0 \}$. The reason will become clear in a moment when we define the Wasserstein distance for discrete data sets. We  define the persistence diagram as the collection of points $\{a_1,\cdots,a_n\} \cup \Delta$, where $a_i \in \R_{\Delta}^2$.

We are interested in the distance between two persistence diagrams, $X$ and $Y$. The metric used is a discrete analog of the more general Wasserstein metric~\cite{Villani2009}. 
We are tasked with matching points from opposing diagrams most efficiently so as to minimize the work\footnote{In the general case, the mass is variable, so transport between distributions involves the traditional {\em work = mass $\times$ distance} formulation. We still use this terminology, except {\em mass} = 1, so we neglect it.} necessary to transport the configuration of points in $X$ to match the configuration of points in $Y$. In the case of discrete mass, we model the $p$-Wasserstein distance as~\cite{Berwald2018a}
\begin{align}\label{eq:wass-dfn}
	d_p(X,Y) = \inf_{\phi:X \to Y} \left( \sum_{a \in X} \|a - \phi(a)\|_q^p \right)^{1/p},
\end{align}
where the infimum is taken over all bijections $\phi$ between points in diagrams $X$ and $Y$ and $p,q \in [1, \infty)$. It is convenient to use $p=q=2$.
Given a specific $\phi$, define the {\em cost} of the matching as 
\begin{align}\label{eq:cost}
	C_p(X,Y) = \sum_{a \in X} \|a - \phi(a)\|_q^p,
\end{align}
where we omit reference to $\phi$ on the left-hand side.

In practice, the problem is often solved by translating the bijection problem to a matching problem on a bipartite graph. Suppose $X$ and $Y$ are persistence diagrams with $m$ and $n$ points, respectively.
We convert the off-diagonal points in $X$ and $Y$ to a weighted bipartite graph representation. Define a diagonal pairing, $\Delta_a$, to be the projection of $a \in \R_{\Delta}^2$ onto its nearest point in the diagonal, under the $\ell_{\infty}$ norm. Set $X_{\Delta} = \{\Delta_{a_i} \mid a_i \in X \}$ and $Y_{\Delta} = \{\Delta_{b_i} \mid b_i \in Y \}$. We now specify the \wg{} graph used to construct a QUBO that we embed on the D-Wave QPU.

\begin{dfn}\label{def:wassgraph}
	Let $\bar{X} := X \cup Y_{\Delta}$ and $\bar{Y} := Y \cup X_{\Delta}$.  Define the \wg{} $W := \langle{\bar{X} \cup \bar{Y}, E}\rangle$, with the edges $E := E_1 \cup E_2 \cup E_3$ such that 
	\begin{align*}
	E_1 =& \{(u,v, \theta_{u,v}) \mid u \in X, v \in Y\} \\
	E_2 =& \{(u,\Delta_u, \theta_{u,\Delta_u}) \mid u \in X, \Delta_u \in X_{\Delta}\} \\
	E_3 =& \{(v,\Delta_v, \theta_{v, \Delta_v}) \mid v \in Y, \Delta_v \in Y_{\Delta}\},
	\end{align*} 
	where the edge weights are defined by 

	\begin{equation}
		\theta_{u,v} = 
		\begin{cases}
			\| u-v \|_{\infty} & \text{if } v = \Delta_u \\
			\| u-v \|_{p} & 	{otherwise}.
		\end{cases}
	\end{equation}
\end{dfn}

\begin{figure}[t]
	\centering
	\includegraphics[width=0.7\textwidth]{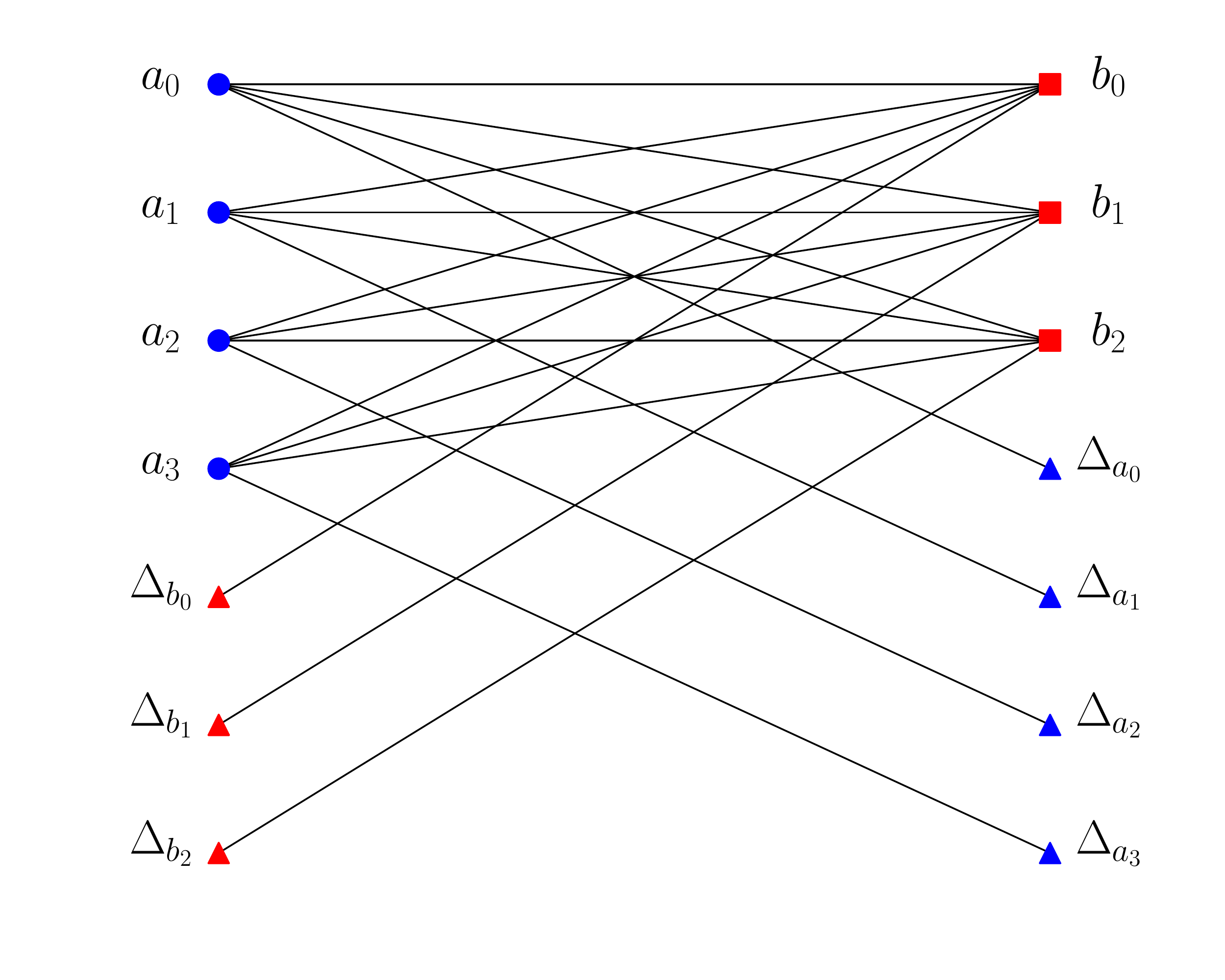}
	\caption{Bipartite graph with nodes from $\bar{X} = \{a_0,\ldots, a_3\} \cup \{\Delta_{b_0}, \Delta_{b_1}, \Delta_{b_2}\}$ labeled by \coloredcircle[blue, fill=blue]{3pt}'s and ${\color{red}\blacktriangle}$'s, 
	and nodes from $\bar{Y} = \{b_0,\ldots,b_2 \} \cup \{\Delta_{a_0}, \ldots, \Delta_{a_3}\}$ labeled by \coloredbox{red} and ${\color{blue}\blacktriangle}$'s. The graph is complete only among off-diagonal points.}
	\label{fig:wass-graph-example}
\end{figure}

\cref{fig:wass-graph-example} shows an example \wg{} for two persistence diagrams. In this case, $X$ contains four points and $Y$ three points. Only the subgraph containing off-diagonal points in $X$ and $Y$ is complete. The complexity of the discrete problem can be reduced significantly in both the classical and quantum computing cases by limiting the number of possible bijections. It is especially beneficial in quantum situations where qubits are at a premium. 

\subsection{The \wg{} as a QUBO}
\label{sec:wg-qubo}

The \wg{} provides a succinct example of how one might bridge mathematics and quantum computers. Given two persistence diagrams, we construct a QUBO from the associated \wg{}, $W$. The approach is straightforward. The QUBO must encode an objective function which minimizes the work, $C_p$, by ``turning on'' specific edges, while also enforcing certain constraints. 

To make this precise, first we enumerate the edges that will map to the logical qubits. The number of edges in $W$ is $N = mn + m +n$, where $m = |X|$ and $n=|Y|$. 
The weighted edges in $W$ map to a set of tuples, 
$\{(x_i, \theta_i) \mid x_i \ \in \Z_2 \}$, that is in one-to-one correspondence with the elements of $E$ so that each edge $(u,v, \theta_{u,v})$ maps to exactly one $(x_i, \theta_i)$. An edge is {\em activated} if $x_i = 1$, otherwise it is {\em inactivated}. We now rewrite the cost in terms that include the logical qubits,
\begin{align*}
	H_{\text{cost}}(\x) = \sum_{(u,v) \in E} \theta_{u,v} x_{u,v} 
\end{align*}
where $\x = (x_{u,v}) \in \B^N$.

To avoid the case where setting $\x = {\bf 0}$ minimizes the objective, we must add constraints. Each $u \in X \cup Y$ must have degree one to avoid duplication of mass and to assure that the points are transported between diagrams. The diagonal nodes can have degree zero or one, depending on whether or not their off-diagonal partner connects to another off-diagonal node. From these requirements we obtain
\begin{align*}
	H_{\text{constraint}}(\x) = \sum_{u \in X} \left(1 - \underset{(u,v) \in E}{\sum_{v \in Y}} x_{u,v}\right)^2 
			 + \sum_{v \in Y} \left(1 - \underset{(u,v) \in E}{\sum_{u \in X}} x_{u,v}\right)^2,
\end{align*}
where we consider only edges emanating from nodes in $X \cup Y$. The summand, $(1- *)$ enforces the requirement that off-diagonal nodes have degree one. If each node in $X \cup Y$ has degree one, then $H_{\text{constraint}} = 0$; otherwise, one or both of the terms in the expression will be positive and add a penalty to the objective function. 

Combining $H_{\text{cost}}$ and $H_{\text{constraint}}$, we arrive back at the general definition of the problem in \cref{eq:minH} with
\begin{align}
\label{eq:wass-qubo}
	H_P(\x) = H_{\text{cost}} + \gamma H_{\text{constraint}}, 
\end{align}
where we have inserted the Lagrangian parameter $\gamma$ to balance the magnitude of the terms. Quantum computers are analog physical devices that have limited accuracy and ranges for their parameters. Thus, determining correct parameters is essential for accurate solutions. 

\begin{figure}[t!b]
	\centering
	\includegraphics[width=0.7\textwidth]{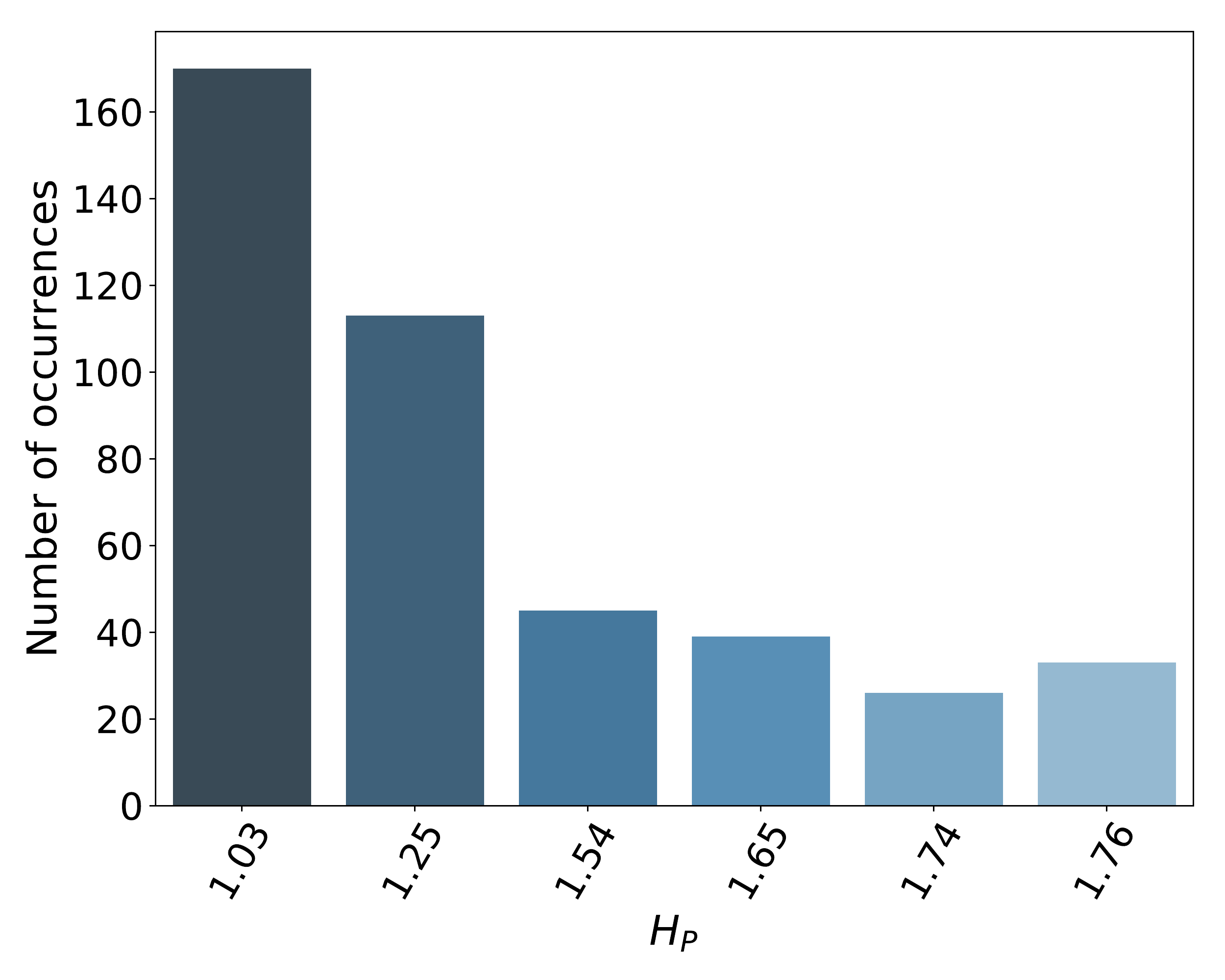}
	\caption{Frequency of costs, $H_P$, computed on the D-Wave quantum computer for different matchings between two persistence diagrams. The diagrams were computed from a torus and an annulus. For this example, $\gamma = 1$. Each distance on the $x$-axis corresponds to a specific $\x \in \B^N$. The occurrences on the $y$-axis denote the number of times a unique $\x$ was sampled. The matching corresponding to the Wasserstein distance, $1.03$ is sampled most frequently.}
	\label{fig:samples}
\end{figure}

\cref{eq:wass-qubo} contains an important question. While the determination of the objective function and its constituent constraints is straightforward, it is not entirely clear that \cref{eq:wass-qubo} yields the same value as \cref{eq:wass-dfn}.  
In \cite[Sec. 4]{Berwald2018a}, we prove that \cref{eq:wass-qubo} computes the $p$-Wasserstein distance, with the caveat that the minimizer of $\HH$ provides an equivalent solution to \cref{eq:wass-dfn} iff $B$ satisfies

\begin{align*}
	B > \max_{(u,v) \in E} \theta_{u,v}.
\end{align*}

By keying the analysis of the quantum computational problem off of a known computable metric, we are able to determine exactly how to set hyperparameters properly. By contrast, it is often necessary in general problems to perform expensive searches of the parameter space before a reasonable energy landscape, defined by $H_P$, can be processed accurately by the QPU.

\subsection{Sampling solutions}
\label{sec:samples}

Quantum computers are inherently probabilistic. Hence, it is necessary to sample the energy landscape of the problem many times to obtain a distribution of solutions. For example, Shor's algorithm~\cite{Shor1995} is designed to return the prime factors of a number with {\em high probability}. Repeated sampling will provide many factor pairs, leaving the final confirmation up to a classical computer. 

On the D-Wave quantum computer, once the energy landscape is defined by $H$, and the QUBO is embedded on the QPU, a collection of hundreds or thousands of samples, $\{ \x_i \}$, can be gathered quickly. The $\x_i$ may be nearby local minima, using Hamming distance as a metric, or one may find solutions with similar energy at opposite corners of the hypercube $\B^N$. 

In the example in \cref{fig:samples}, the suite of samples we gather represents different possible matchings between the persistence diagrams and their associated cost, $C_P$. We use $H_P(\x)$ to compute the cost. The low-energy solutions  represent valid matchings that do not violate constraints, eg., $H_{\text{constraint}}(\x)=0$. The minimum cost, $1.03$, is the square of the Wasserstein distance, i.e., the infimum over all the possible valid matchings. 

The different matchings and distances represent a distribution of low-energy solutions, each of which comes from a different choice of $\phi$ and produces a different cost using \cref{eq:cost}. In fact, \cref{fig:samples} represents a distribution of $\phi$'s sampled from an approximate Boltzmann distribution. 
In future work we plan to study the implications for statistics on persistence diagrams, along the lines outlined by Turner \etal., in their work on Fr\'{e}chet means in \cite{Turner2014}.



\section{Conclusion}
\label{sec:concl}

In this article we covered a number of mathematical aspects of quantum computing from a high level. Nevertheless, we have hardly scratched the surface of the subject. Interesting problems can be found in many different areas, from physical applications, to theoretical improvement of embedding QUBOs on the QPU, to decomposition of large problems into QPU-sized chunks. 

Mathematicians and physicists have spent many years developing algorithms designed to run faster on quantum computers. The subtlety is that many of these methods require far more qubits than are available even on the 2000-qubit D-Wave quantum computer. Luckily, even before we reach that technological state, there is still exciting and effective research that can be accomplished in the current NISQ era. 
We hope that in touching on the mathematics involved in programming a D-Wave quantum computer we motivate interest in the myriad problems stemming from using this novel computational tool.


\paragraph{Acknowledgements:}
 The author gratefully acknowledges support from the Institute for Mathematics and its Applications at the University of Minnesota. Many thanks are also due to Joel Gottlieb, Elizabeth Munch, and Steve Reinhardt for helpful discussions and suggestions.

\bibliographystyle{ieeetr}
\bibliography{QuantumBib}

\end{document}